\documentclass[12pt,superscriptaddress,showpacs,showkeys,twocolumn,prb,aps,reprint,a4paper,floatfix]{revtex4-2}

\usepackage{graphicx}
\usepackage{amsmath, amsthm, amssymb, amsfonts}
\usepackage{float}

\usepackage{natbib}
\usepackage{physics}
\usepackage[para,online,flushleft]{threeparttable} 
\usepackage{array}
\usepackage{ragged2e} 

\makeatletter
\newcommand*{\citenst}[2][]{%
  \begingroup
  \let\NAT@mbox=\mbox
  \let\@cite\NAT@citenum
  \let\NAT@space\NAT@spacechar
  \let\NAT@super@kern\relax
  \renewcommand\NAT@open{[}%
  \renewcommand\NAT@close{]}%
  \citet[#1]{#2}%
  \endgroup
}
\makeatother

\makeatletter
\newcommand*{\citenumns}[2][]{%
  \begingroup
  \let\NAT@mbox=\mbox
  \let\@cite\NAT@citenum
  \let\NAT@space\NAT@spacechar
  \let\NAT@super@kern\relax
  \renewcommand\NAT@open{[}
  \renewcommand\NAT@close{]}%
  \cite[#1]{#2}
  \endgroup
}
\makeatother
\usepackage[caption=false]{subfig}

\usepackage{lipsum}

\begin{document}
\title{Superconducting NbN Resonator Parametric Amplifiers for Millimetre Wavelengths}

\author{Songyuan Zhao}
\email{Author to whom any correspondence should be addressed.\\ E-mail: songyuan.zhao@physics.ox.ac.uk}
\affiliation{Clarendon Laboratory, Parks Road, Oxford OX1 3PU, United Kingdom.}
\author{S. Withington}
\affiliation{Clarendon Laboratory, Parks Road, Oxford OX1 3PU, United Kingdom.}
\author{C. N. Thomas}
\affiliation{Cavendish Laboratory, JJ Thomson Avenue, Cambridge CB3 OHE, United Kingdom.}
\date{\today}

\begin{abstract}
\noindent We report the development of a reactive sputtering process for high $T_\mathrm{c}$ NbN films with high normal-state resistivity, tailored for kinetic inductance parametric amplifiers. The process includes precise control to ensure full nitridation of the target prior to deposition. Under optimized conditions, the resulting NbN thin films exhibit a critical temperature of $10.5\,\mathrm{K}$ and a resistivity of $\sim1000\,\mathrm{\mu\Omega\,cm}$. The high $T_\mathrm{c}$ of the NbN thin-films suggests strong potential for application over the entire millimetre-wave frequency range from $24\,\mathrm{GHz}$ to $300\,\mathrm{GHz}$, whereas the high resistivity suggests a reduced power requirement for the pump tone to achieve high gain. Resonator parametric amplifiers have been fabricated from these films using coplanar waveguide geometry. The devices were able to produce high gain exceeding $20\,\mathrm{dB}$ at $25\,\mathrm{GHz}$, with artefact-free, reproducible amplification profiles in good agreement with theoretical models.
\end{abstract}


\maketitle

\section{Introduction}

Superconducting parametric amplifiers have undergone extensive development in recent years and are now widely employed for ultra-low-noise readout in quantum information, quantum sensing, observational physics, and fundamental physics experiments \citenumns{Ranzani_2018, Zobrist_2019, Vissers_2020, Project8_2009, Saakyan_2020, QTNM_collaboration_white_paper}. These amplifiers can operate with added noise approaching the Standard Quantum Limit (SQL), offering nearly an order of magnitude improvement over conventional semiconductor-based high electron mobility transistor (HEMT) amplifiers \citenumns{Eom_2012,McCulloch_2017}. Two primary approaches are employed to realise such devices: using the nonlinear inductance of Josephson junctions, as in Josephson Parametric Amplifiers (JPAs), or using the nonlinear kinetic inductance of superconducting thin-films, as in Kinetic Inductance Travelling-Wave Parametric Amplifiers (KI-TWPAs) and Kinetic Inductance Resonator Parametric Amplifiers (KI-ResPAs). Amplifiers based on both types of nonlinearities have demonstrated added noise close to the SQL \citenumns{Yurke_1988,Malnou_2020}.

In this paper, we focus on parametric amplifiers based on the nonlinear kinetic inductance, which has lower fabrication requirements (e.g. single layer coplanar waveguide with no sub-$\mu m$ features), and can be designed to have greater saturation powers, higher maximum operating frequencies, and higher operating temperatures \citenumns{Zhao_2023,zhao2024_intrinsic_separation}. To date, JPAs typically operate below $10\,\mathrm{GHz}$ \citenumns{JPA_f_range,Mutus_JPA_range,JPA_f_range_ANR}. As we demonstrate in this study, kinetic inductance parametric amplifiers based on NbN thin-films can operate at least up to $25\,\mathrm{GHz}$, and likely across the entire millimetre-wave frequency range.

Both KI-TWPAs and KI-ResPAs are capable of achieving high gain exceeding $20\,\mathrm{dB}$ with added noise close to the SQL \citenumns{Macklin_2015,Eom_2012,Tholen_2009}. While the NbN thin-films developed in this study are suitable for use in both geometries, we focus here on resonator-based parametric amplifiers (ResPAs). ResPAs are narrow-band devices, typically offering bandwidths ranging from several to hundreds of $\mathrm{MHz}$. They have lower pump power requirements and are less prone to lithographic defects due to their short physical lengths (e.g. $0.1-1\,\mathrm{cm}$) \citenumns{Shan_2016,Tholen_2009,Zhao_2023}. The ease-of-fabrication of ResPAs means tens to hundreds of amplifiers can be fabricated on a single wafer whilst maintaining extremely high yield. This technology is ideally suitable for narrow-band applications that require high gain, low noise amplification, especially in the format of large arrays. These narrow-band applications span fundamental physics experiments, including direct measurement of neutrino mass through cyclotron radiation emission spectroscopy \citenumns{Oblath_2020,Saakyan_2020,QTNM_collaboration_white_paper}, dark matter searches \citenumns{Axion_2024,HiddenSector_2024,Adams_2022_Snowmass}, and readout of detectors for astronomy \citenumns{HAMILTON_1980,Westig_2018}, as well as quantum computing and control, including readout of qubits \citenumns{Devoret_2013,Naaman_2022}, quantum feedback \citenumns{Vijay2012_feedback}, quantum error detection \citenumns{Córcoles_2015,Riste_2015}, and measurement of quantum nanomechanical oscillators and nanobolometers \citenumns{CLeland_2002,Teufel_2011,Kokkoniemi_2019}. In particular, the measurement of neutrino mass requires large-format inward-looking phased arrays to enable detection volumes of up to several cubic metres. These phased arrays in turn necessitate large numbers of quantum-noise-limited amplifiers to achieve the required sensitivity \citenumns{withington2024quantum,QTNM_collaboration_white_paper}. 

Over the past several years, there has been growing interest in kinetic inductance parametric amplifiers in the millimetre-wave regime, motivated by the need for ultra-low-noise amplification in emerging high-frequency quantum computing platforms \citenumns{Anferov_2024_20GHz_Qubits,Anferov_2025_mmWave_Qubits,Zhao_2026_scalable}, millimetre-wave astronomy receivers \citenumns{Belitsky_2018_ALMA_Band5,Tan_2024_mmWave_KITWPA}, and fundamental physics experiments \citenumns{QTNM_collaboration_white_paper, Project8_2023_expt}. Several design studies have explored the potential of millimetre-wave parametric amplifiers \citenumns{Tan_2024_mmWave_KITWPA,Mena_2024_mmWave_CPW,Banys_2022_millimetre}. A direct experimental realisation was demonstrated by Shu \textit{et al.}, who reported a NbTiN travelling-wave parametric amplifier with above $10\,\mathrm{dB}$ of gain over a broad frequency range from $3$ to $34\,\mathrm{GHz}$ \citenumns{Shu_2021}.


The performance of a kinetic inductance parametric amplifier is strongly determined by the properties of the underlying material, as parametric wave-mixing is enabled by the distributed nonlinear kinetic inductance of the thin-film \citenumns{Eom_2012,zhao2022physics}. From an application perspective, the superconducting transition temperature $T_\mathrm{c}$ and the normal-state film resistivity are particularly important parameters: 

The $T_\mathrm{c}$ sets a limit on the operating temperature $T$. Not only does $T$ have to be smaller than $T_\mathrm{c}$, in superconducting detectors such as Kinetic Inductance Detectors, the ratio of $T$ to $T_\mathrm{c}$ is typically kept below $1/5$ in order to ensure saturation in the quasiparticle lifetime and reduce the recombination noise \citenumns{Barends_Lifetime_2008,Jonas_review}. Importantly, $T_\mathrm{c}$ also sets a limit on the operating \textit{frequency} of the parametric amplifiers. For BCS superconductors as well as many unconventional superconductors, the superconducting energy gap $\Delta_\mathrm{g}$ scales according to the $T_\mathrm{c}$, and the relation is given by $\Delta_\mathrm{g}=1.76\,k_\mathrm{B}T_\mathrm{c}$ in the BCS theory. Direct time-resolved spectroscopy on superconducting NbN reported a slightly larger gap ratio for NbN thin films, with $\Delta_\mathrm{g}\sim2.3\,k_\mathrm{B}T_\mathrm{c}$ \citenumns{Beck_2011_NbN_gap}. Electromagnetic waves with energy greater than \textit{twice} the energy gap will break Cooper pairs into quasiparticles and disrupt the superconducting state. For a BCS superconducting material to retain superconductivity across the entire millimetre-wave range, the $T_\mathrm{c}$ needs to be greater than $4\,\mathrm{K}$. While this $T_\mathrm{c}$ requirement is high for elemental superconductors, it can be achieved using disordered BCS superconductors like NbTiN, TiN, and NbN.

The resistivity of the superconducting thin-film is also important to kinetic inductance parametric amplifiers. The kinetic inductance of a superconducting transmission line has the following form, up to second order in the current $I$:
\begin{align}
L_{\mathrm{k}} = L_{\mathrm{k},0}\left[1+\left(\frac{I}{I_*}\right)^2\right] \, , \label{eq:nonlinear_scale}
\end{align}
where $L_{\mathrm{k}}$ is the kinetic inductance per unit length, $L_{\mathrm{k},0}$ is the kinetic inductance per unit length in the absence of inductive nonlinearity, and $I_*$ characterises the scale of the nonlinearity. The total inductance per unit length $L$ comprises contributions from both the kinetic inductance $L_{\mathrm{k}}$ and the geometric inductance $L_{\mathrm{g}}$. Parametric wave-mixing occurs when the current on the transmission line becomes significant compared to $I_*$, which is comparable to the critical current $I_\mathrm{c}$ \citenumns{Eom_2012,Jonas_review,zhao2022physics} and is likewise reduced when the normal-state resistivity increases. Further, a thin-film with high normal-state resistivity will also benefit from a higher ratio of the nonlinear kinetic inductance to the linear geometric inductance \citenumns{Jonas_review,zhao2022physics}, likewise enhancing the effect of parametric wave-mixing. In total, these effects mean that a higher normal-state resistivity reduces the pump power required to achieve high gain parametric amplification. This is highly beneficial because stronger pump powers can introduce unwanted power-handling instabilities and increase the thermal load on the device, packaging, and cryogenic environment, which can be particularly important when the amplifier is operated at sub-Kelvin temperatures. Furthermore, a stronger pump carrier produces a larger phase-noise skirt, whose power spectrum scales with the pump power and may degrade signal fidelity when the signal frequency is close to the pump or its harmonics \citenumns{shiri_2026_phaseNoise}.

In this paper, we report the development of a reactive sputtering process for high $T_\mathrm{c}$ NbN films with high resistivity, tailored for use in kinetic inductance parametric amplifiers. Under optimized conditions, the resulting NbN thin films exhibit a critical temperature of $10.5\,\mathrm{K}$ and a resistivity of $\sim1000\,\mathrm{\mu\Omega\,cm}$. The high $T_\mathrm{c}$ of the NbN thin-films suggests strong potential for application over the entire millimetre-wave frequency range, whereas the high resistivity suggests a reduced power requirement for the pump tone to achieve high gain. Resonator parametric amplifiers were fabricated from these films using coplanar waveguide geometry. The devices were able to produce high gain exceeding $20\,\mathrm{dB}$ at $25\,\mathrm{GHz}$, with artefact-free, reproducible amplification profiles in good agreement with theoretical models. The measurement frequency of $25\,\mathrm{GHz}$ was due to the limit of SMA connectors in the readout electronics, and indeed the artefact-free, reproducible amplification profiles, as well as the high $T_\mathrm{c}$ suggest that thin-film NbN will be suitable for high frequency operation in the millimetre-wave range. 

\section{NbN Deposition Process}
\begin{figure}[htb!]
\includegraphics[width=8.6cm]{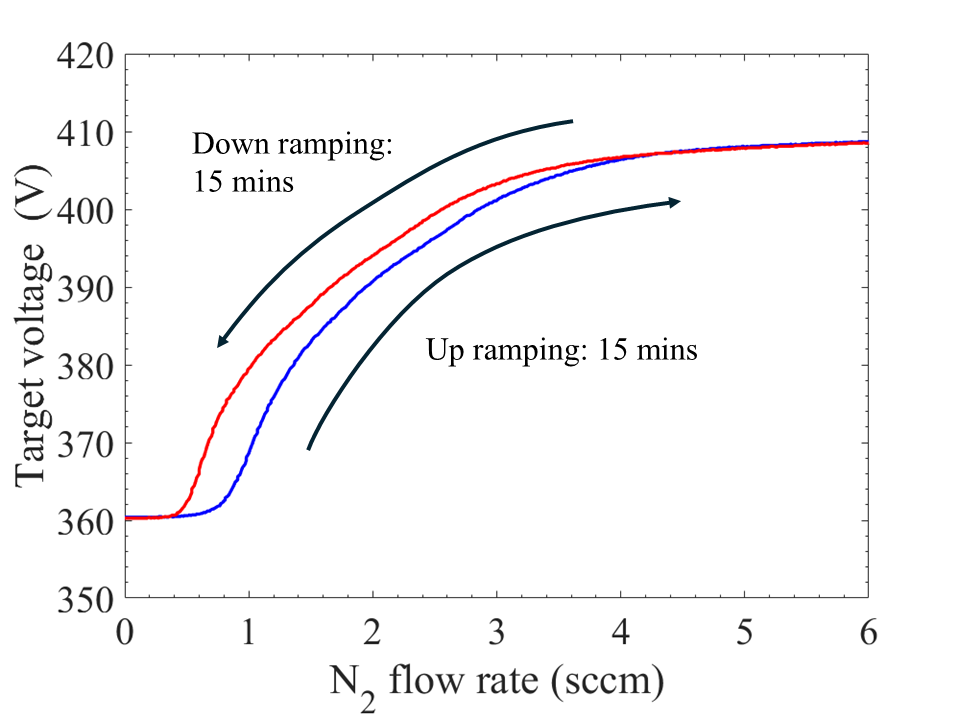}
\caption{\label{fig:1_hysteresis} Target voltage as a function of nitrogen flow in the presence of a constant argon 
flow of 14.0 sccm and power of 100 W. The blue plot shows the up ramping of nitrogen flow over 15 minutes until the Nb target became saturated; the red plot shows the down ramping of nitrogen flow over 15 minutes after saturation had occurred. The difference in target voltage between the two plots illustrates target hysteresis.}
\end{figure}

\begin{figure}
\includegraphics[width=8.6cm]{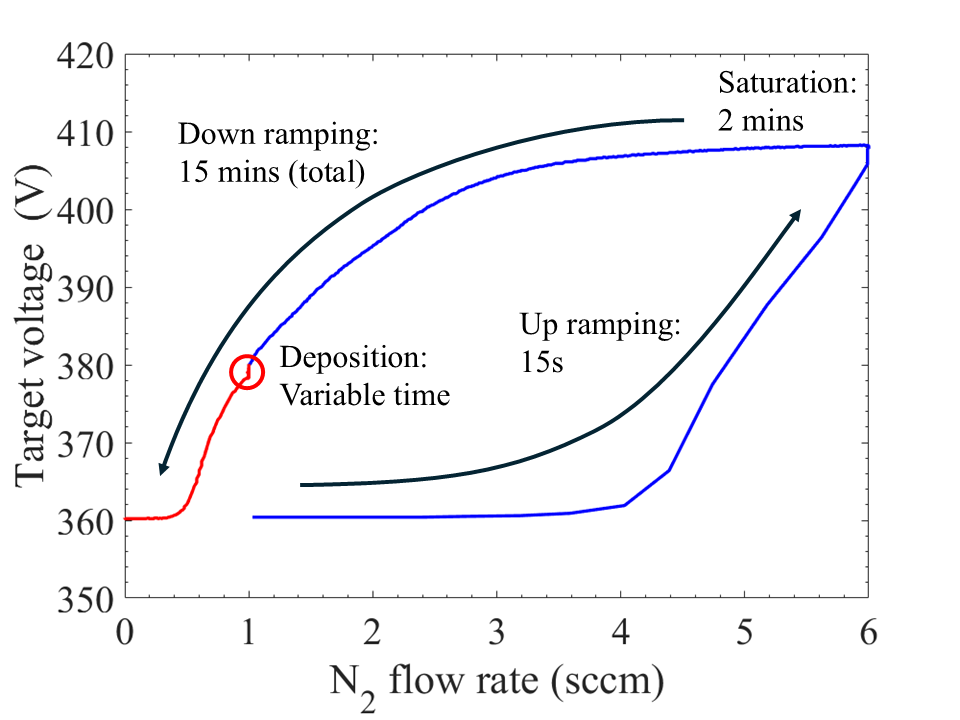}
\caption{\label{fig:2_ramping} Routine for film deposition. Target voltage as a function of nitrogen flow in the presence of a constant argon flow of 14.0 sccm and power of 100 W. Nitrogen flow was rapidly increased to 6.0 sccm in 15 seconds and held for 2 minutes to saturate the Nb target, after which it was gradually reduced to the chosen flow rate for deposition. Deposition time was dependent on the desired film thickness; for example, a $100 \,\mathrm{nm}$ NbN film required approximately 28 minutes. Following deposition, the nitrogen flow continued to ramp down, with a total down-ramping duration of 15 minutes.}
\end{figure}

Deposition was carried out using an ultra-high vacuum DC magnetron sputtering system with a base pressure at or below $2 \times 10^{-10}$ Torr. The process gas for the deposition was argon. The argon flow rate was set to $14.0\,\mathrm{sccm}$, corresponding to a partial pressure of $2.7\,\mathrm{mTorr}$, and the target power was kept constant at $100\,\mathrm{W}$. The NbN films were deposited by reactive sputtering of a 4-inch Nb target with a manufacturer-rated purity of $99.95\%$ in the presence of gaseous $N_2$ onto $7.5\,\mathrm{mm} \times 13.5\,\mathrm{mm}$ RCA-cleaned silicon substrates. For deposition at an $N_2$ flow rate of $1.0\,\mathrm{sccm}$, the chamber pressure was measured to be $2.9\,\mathrm{mTorr}$. The sputtering voltages corresponding to the deposition conditions are shown in figure~\ref{fig:1_hysteresis} and figure~\ref{fig:2_ramping}. No substrate heating was applied during deposition, and all depositions were performed at room temperature. The target was first pre-sputtered for 5 minutes before each deposition run to remove surface contamination from the Nb target.

The properties of NbN films are strongly dependent on the nitrogen flow rate. As discussed in the introduction, for applications in kinetic inductance parametric amplifiers, the films should have high $T_\mathrm{c}$ and high resistivity. In order to find suitable deposition conditions, we first measured the hysteresis curve of target voltage against nitrogen flow. 

As shown in figure~\ref{fig:1_hysteresis}, the target voltage against nitrogen flow was hysteretic and took on different values depending on whether the nitrogen flow was ramping up \textit{toward target saturation} or ramping down \textit{after target saturation}. Target saturation here refers to the condition where the Nb target surface becomes fully poisoned, forming a nitride layer on the target face. Both the ramp-up and ramp-down processes were computer-controlled and carried out continuously over 15 minutes without discrete steps. For our deposition system, we observed that saturation occurred at around $5\,\mathrm{sccm}$, where the two branches of hysteresis converged. Previous studies have shown that NbN films with high $T_\mathrm{c}$ can be deposited at the pump-down branch of the hysteresis curve around the region where the target voltage varies most rapidly against nitrogen flow, and where the difference between the two branches is the largest \citenumns{NbN_hysteresis,glowacka_NbN}. This was measured at a nitrogen flow of approximately $1 \,\mathrm{sccm}$ for our deposition environment. This region served as an approximate nitrogen flow range around which films were deposited and characterised to identify the flow that gave the desired NbN thin film properties.

The deposition conditions were further fine-tuned by depositing NbN thin-films at a nitrogen flow of around $1 \,\mathrm{sccm}$ and measuring their properties. In these depositions, instead of slowly ramping up to saturation over 15 minutes, we performed a quick ramping to the saturation point over 15 seconds. The nitrogen flow was then held at $6\,\mathrm{sccm}$ for 2 minutes to establish target saturation. Afterwards, the flow was ramped down to the desired flow rate and the shutter was opened for film deposition. The time taken for film deposition was dependent on the desired film thickness, which was checked using a Dektak profilometer. As a reference, a $100 \,\mathrm{nm}$ NbN film required approximately 28 minutes of deposition time. An example of this deposition routine is shown in figure~\ref{fig:2_ramping}.

\begin{figure}
\includegraphics[width=8.6cm]{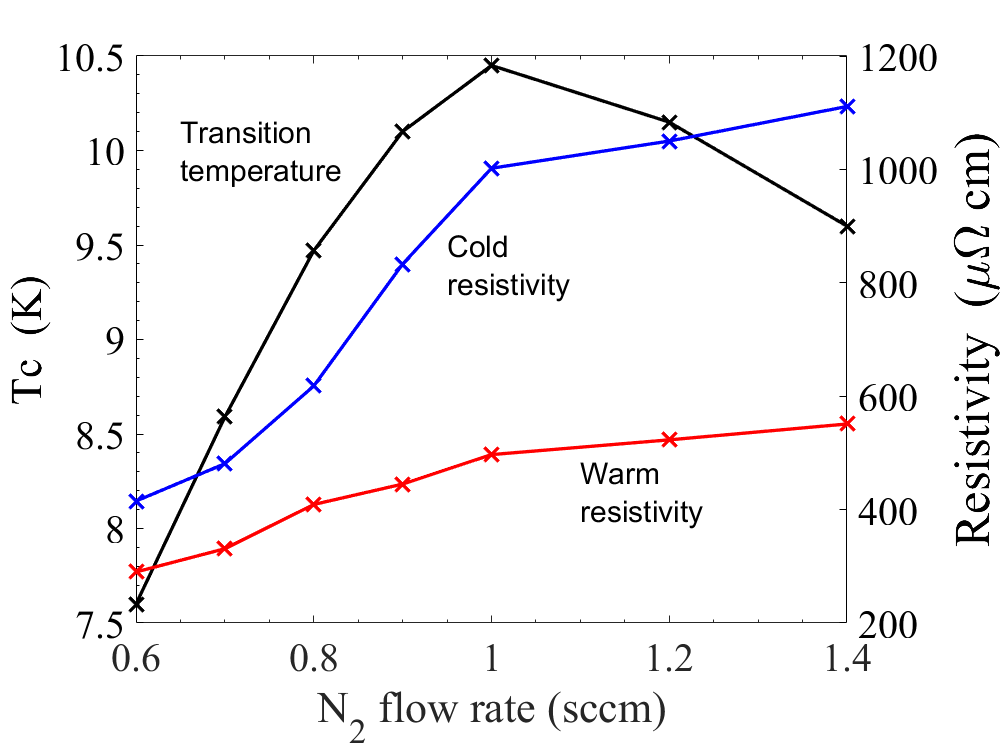}
\caption{\label{fig:3_properties} Properties of NbN films deposited at different nitrogen flow rates. Black markers: transition temperature; blue markers: cold resistivity just above the superconducting transition; red markers: warm resistivity at room temperature.}
\end{figure}

The thin-films were configured for four-wire resistance measurement using an AVS Resistance Bridge. The films were placed in a custom enclosure and mounted at the cold stage of our adiabatic demagnetization refrigerator. The temperature of the enclosure was monitored using a calibrated ruthenium oxide thermometer and the refrigerator was cooled down and warmed up repeatedly to establish the transition temperature of the NbN thin films. The film resistivity was measured at room temperature (warm resistivity) and just above the superconducting transition (cold resistivity).

The measured properties of $100\,\mathrm{nm}$ thin-films deposited in the nitrogen flow range of $0.6-1.4\,\mathrm{sccm}$ are shown in figure~\ref{fig:3_properties}. As seen in the figure, the transition temperature peaked at $10.5\,\mathrm{K}$, corresponding to the thin-film deposited at $1.0\,\mathrm{sccm}$. For a first-pass estimation using BCS theory, the pair-breaking frequency $f_{2\Delta_\mathrm{g}}\sim770\mathrm{GHz}$, where $k_\mathrm{B}$ is the Boltzmann constant and $h$ is the Planck constant. This suggests that NbN could be used for quantum electronics over the entire millimetre-wave frequency band from $24\,\mathrm{GHz}$ to $300\,\mathrm{GHz}$. Both warm and cold resistivity increased with nitrogen flow. The resistivity achieved was higher compared to previous studies, e.g., our films were $\sim10\,\%$ more resistive compared to those in \citenumns{NbN_hysteresis} and almost twice as resistive compared to \citenumns{glowacka_NbN}. This is likely because these previous studies prioritised optimising $T_\mathrm{c}$ instead of balancing $T_\mathrm{c}$ with resistivity. The $T_\mathrm{c}$ of our NbN films can be further increased through optimisation of the argon flow rate, as demonstrated in \citenumns{NbN_high_Tc,NbN_hysteresis}. More broadly, both $T_\mathrm{c}$ and resistivity could be further optimised over a wider parameter space, including the sputtering current and voltage, substrate temperature, argon-to-nitrogen ratio, and chamber pressure. Such optimisation lies beyond the scope of the present study. Our target application is millimetre-wave parametric amplification at sub-Kelvin temperature, and a resistive NbN film with $T_\mathrm{c}$ above $10\,\mathrm{K}$ is sufficient.

\section{Parametric Amplification at 25 GHz}
\begin{figure}[htb!]
\includegraphics[width=8.6cm]{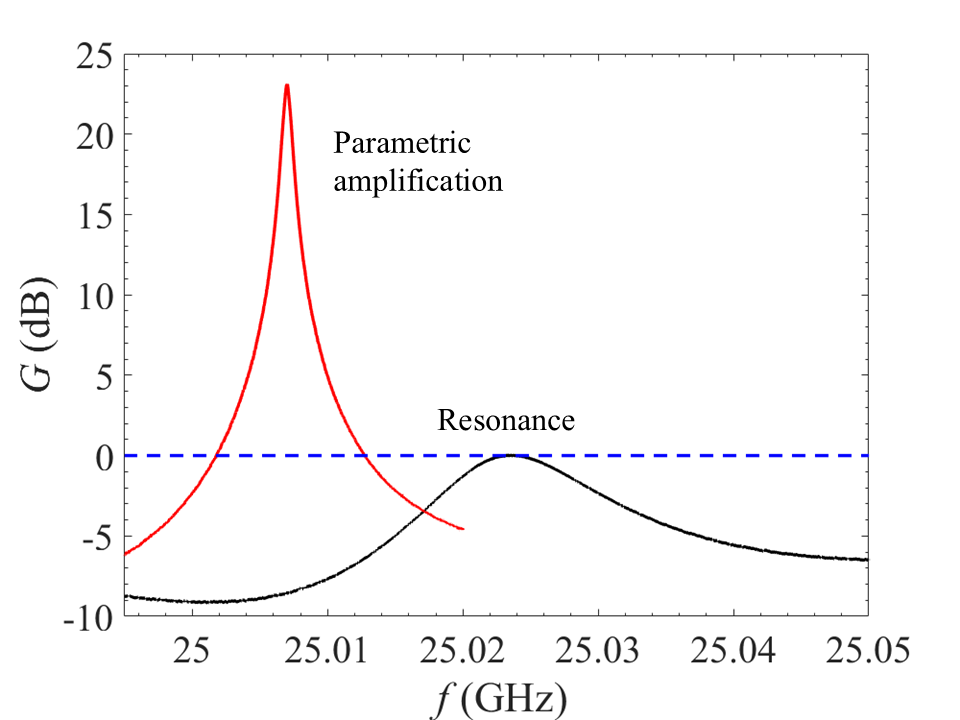}
\caption{\label{fig:4_amplification} Resonance transmission and parametric gain of the NbN resonator amplifier deposited at $1.0\,\mathrm{sccm}$ operating at $25\,\mathrm{GHz}$. Black line: response of the resonance in the absence of the pump tone; blue dashed line: reference line of unity gain, i.e. $0\,\mathrm{dB}$; red line: response of the resonance in the presence of the pump tone at $25.007\,\mathrm{GHz}$ with the resonator now behaving as a parametric amplifier. The pump power at the input of the amplifier was $-28.60\,\mathrm{dBm}$. }
\end{figure}

\begin{figure*}
\includegraphics[width=16cm]{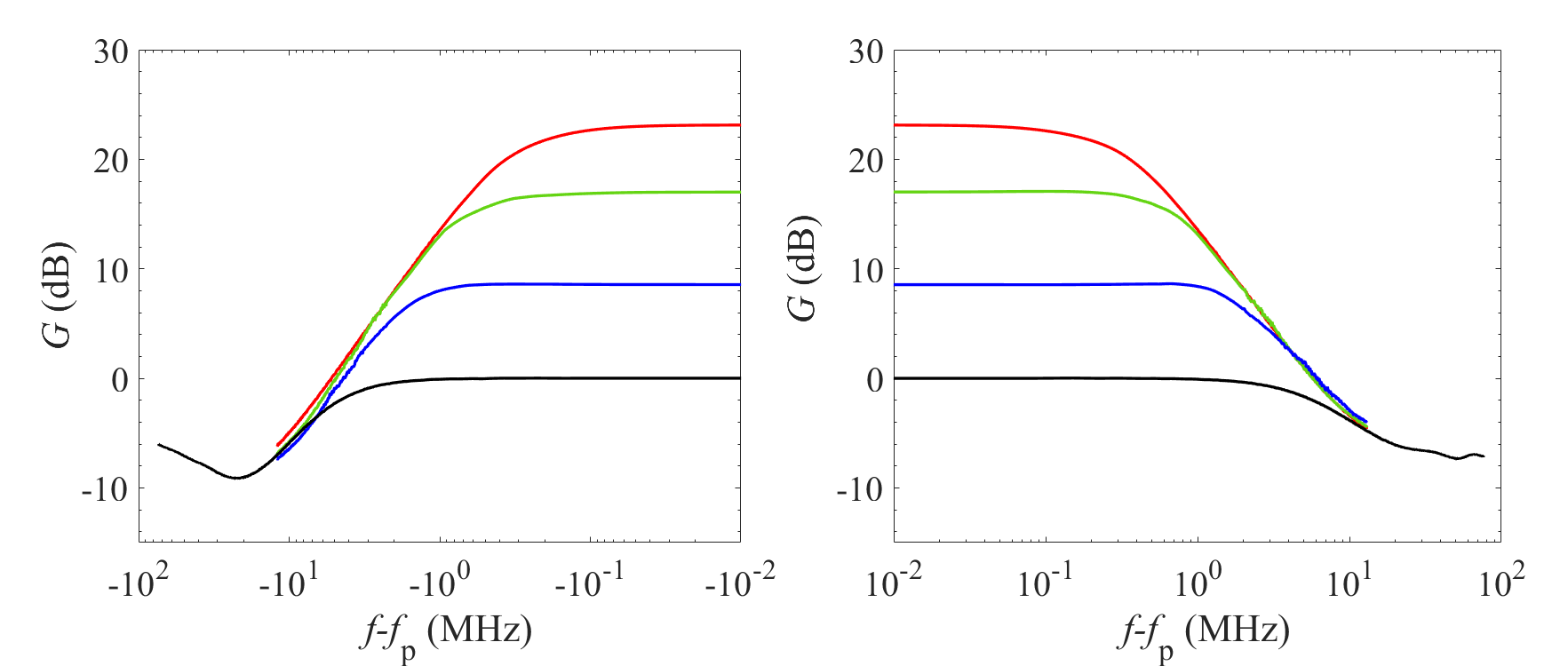}
\caption{\label{fig:5_amplification_log}  Resonance transmission and parametric gain of the NbN resonator amplifier deposited at $1.0\,\mathrm{sccm}$ against the relative frequency $f - f_\mathrm{p}$ on the logarithmic scale. Measurements were performed at $20\,\mathrm{mK}$ in a Bluefors dilution refrigerator. For the black line, $f_\mathrm{p}$ denotes the resonant frequency of $25.023\,\mathrm{GHz}$ in the absence of the pump tone; for the red, green, and blue lines, $f_\mathrm{p}$ denotes the pump frequency of $25.007\,\mathrm{GHz}$. The pump powers at the amplifier input for the red, green, and blue lines were $-28.60\,\mathrm{dBm}$,  $-28.67\,\mathrm{dBm}$, and $-28.74\,\mathrm{dBm}$, respectively.}
\end{figure*}

We fabricated a batch of resonator parametric amplifiers using NbN films deposited using a nitrogen flow of $1.0 \,\mathrm{sccm}$. The general design of the resonator device, its packaging, and its operation as parametric amplifier were discussed in detail in our previous study \citenumns{Zhao_2023}. The amplifier was based on a two-port half-wave resonator housed in a gold-plated copper enclosure with SMA connectors. The resonator had coplanar waveguide geometry with thickness of $100\,\mathrm{nm}$, width of $2\,\mathrm{\mu m}$, gap width of $20\,\mathrm{\mu m}$, and length of $8\,\mathrm{mm}$. The packaged device was mounted at the cold stage of a Bluefors dilution refrigerator whose base temperature was $20\,\mathrm{mK}$. Checks were performed to ensure that the power levels used in this study did not result in unwanted saturation in the readout electronics. Our previous study had demonstrated reliable amplification in the sub-$10\,\mathrm{GHz}$ range. In this study, we operated the resonator parametric amplifier on its eleventh harmonic at $25\,\mathrm{GHz}$. The choice of $25\,\mathrm{GHz}$ was due to the limit of SMA connectors in the device enclosure and the readout electronics, which are reliable up to $26.5\,\mathrm{GHz}$, and should not be taken as a limit for the NbN thin-film. 

Figure~\ref{fig:4_amplification} shows the resonance transmission and parametric gain of the NbN resonator amplifier. In the absence of a pump tone, as shown in the black line, the transmission characteristic was consistent with that of a transmission line resonator \citenumns{Pozar_2011}: full transmission, i.e. $0\,\mathrm{dB}$, occurred on resonance, and a first-order roll-off occurred at detuned frequencies. The $3\,\mathrm{dB}$ bandwidth of the resonance was $13.5\,\mathrm{MHz}$. In the presence of a strong pump tone at $25.007\,\mathrm{GHz}$, the resonator behaved as a parametric amplifier, experimentally demonstrating amplification at $25\,\mathrm{GHz}$, as shown by the red line. A high peak gain of $23\,\mathrm{dB}$ and a $3\,\mathrm{dB}$ bandwidth of $0.7\,\mathrm{MHz}$ were measured. Notably, this high gain was achieved with a low pump power of $-28.60\,\mathrm{dBm}$ at the input of the amplifier. In comparison, the travelling-wave amplifier described in \citenumns{Eom_2012} required a pump power of $-8\,\mathrm{dBm}$ to obtain approximately $10\,\mathrm{dB}$ of gain. The difference in power \textit{density} between the two devices was even more pronounced, given that the travelling-wave amplifier had a film thickness of $35\,\mathrm{nm}$ and a CPW conductor width of $1\,\mathrm{\mu m}$, whereas the resonator amplifier in this study had a thickness of $100\,\mathrm{nm}$ and CPW conductor width of $2\,\mathrm{\mu m}$. This figure-of-merit comparison highlights that resonator-based parametric amplifiers realised using highly resistive thin films can indeed provide high gain, narrow-band amplification with low pump power requirements. 

This reduction in the required pump power arises from two factors. Firstly, for a given power incident on a resonator, the energy stored on the resonator is enhanced by its quality factor \citenumns{Thomas_2020}. As a result, the energy required to achieve nonlinear mixing can be satisfied at a lower incident power. Secondly, the NbN thin-film used in this study had high resistivity of $\sim1000\,\mathrm{\mu\Omega\,cm}$, whereas the NbTiN film used in \citenumns{Eom_2012} had a moderate resistivity of $\sim100\,\mathrm{\mu\Omega\,cm}$. As discussed earlier, a higher normal-state resistivity reduces the critical current and increases the kinetic inductance, thereby lowering the pump power required for nonlinear mixing and parametric amplification.

The peak gain of the parametric amplifier can be tuned by adjusting the combination of pump frequency and power \citenumns{Thomas_2022}. In figure~\ref{fig:5_amplification_log}, we have plotted parametric gain from high to moderate gains and the resonance transmission against the relative frequency $f - f_\mathrm{p}$ on the logarithmic scale. For the black line, $f_\mathrm{p}$ denotes the resonant frequency of $25.023\,\mathrm{GHz}$ in the absence of the pump tone; for the red, green, and blue lines, $f_\mathrm{p}$ denotes the pump frequency of $25.007\,\mathrm{GHz}$ which was kept constant. The pump powers at the amplifier input for the red, green, and blue lines were $-28.60\,\mathrm{dBm}$,  $-28.67\,\mathrm{dBm}$, and $-28.74\,\mathrm{dBm}$, respectively. This type of Bode plot is commonly used in amplifier analysis, and visually illustrates concepts such as peak gain, $3\,\mathrm{dB}$ bandwidth, roll-off, and resonance poles. As seen in the figure, although different pump operating points resulted in different values of peak gains, their roll-off behaviour coincided along a common straight-line decay on each side of the Bode plot, with a value of approximately $20\,\mathrm{dB}$ per decade. This common single-pole roll-off is consistent with theoretical analyses of resonator-based parametric amplifiers \citenumns{Thomas_2022}. In contrast to KI-TWPAs, which exhibit rapid gain fluctuations of about $10\,\mathrm{dB}$ over tens of megahertz \citenumns{Eom_2012}, the gain profiles of NbN KI-ResPAs in figure~\ref{fig:5_amplification_log} are notably artefact-free and agree well with theory.

As seen in the measurements above, the resonator parametric amplifier made using $1.0\,\mathrm{sccm}$ NbN thin-films demonstrated excellent amplification characteristics when measured at $25\,\mathrm{GHz}$. These devices have the strong potential to be valuable for narrowband applications in fundamental physics, such as measurement of neutrino mass \citenumns{QTNM_collaboration_white_paper,Saakyan_2020,Oblath_2020} at $\sim20\,\mathrm{GHz}$, as well as in quantum information systems, such as high-efficiency measurement of qubits \citenumns{Qubits_JPA} and quantum error correction \citenumns{Devoret_2013,Córcoles_2015}, usually performed at frequencies below $10\,\mathrm{GHz}$. Recent research has explored various qubit schemes operating at higher frequencies, typically from $\sim 10\,\mathrm{GHz}$ to $24\,\mathrm{GHz}$, where reliable Josephson Parametric Amplifiers are still in early stages of development. Kinetic inductance parametric amplifiers employing high $T_\mathrm{c}$ NbN films with high operating frequencies could therefore offer a promising solution to the need for ultra-low-noise microwave readout in this regime.

\section{Conclusion}
We developed a reactive sputtering process for producing high $T_\mathrm{c}$ NbN films with high resistivity, suitable for use in kinetic inductance parametric amplifiers. Careful control of the process has been exercised to ensure complete nitridation of the target prior to film deposition. Under our optimized conditions, the resulting NbN thin films exhibit a critical temperature of $10.5\,\mathrm{K}$ and a resistivity of $\sim1000\,\mathrm{\mu\Omega\,cm}$. Using these films, we fabricated resonator parametric amplifiers based on coplanar waveguide geometry. These devices achieved high gain exceeding $20\,\mathrm{dB}$ at $25\,\mathrm{GHz}$, with amplification profiles that are artefact-free, reproducible, and consistent with theoretical predictions \citenumns{Thomas_2020,Thomas_2022}.

The artefact-free amplification profiles at $25\,\mathrm{GHz}$ across various gain levels highlight the potential of these devices for high frequency operation. As discussed in previous sections, NbN ResPAs are highly promising for applications such as the direct measurement of neutrino mass and the readout of high frequency qubits, which exceed the frequency limits of current JPA technology. The ease-of-fabrication of ResPAs also makes them highly suitable for deployment in large-array formats, and we are further investigating the use of NbN ResPAs in phased-array antenna systems. As noted earlier, the high critical temperature of the NbN films at $10.5\,\mathrm{K}$ implies a pair-breaking frequency of $\sim770\,\mathrm{GHz}$, suggesting their suitability for operation across the full millimetre-wave frequency band from $24\,\mathrm{GHz}$ to $300\,\mathrm{GHz}$. Future work should explore packaging strategies compatible with waveguide interfaces to enable practical deployment at higher frequencies. Our previous studies at $6-7\,\mathrm{GHz}$ suggest that ResPAs can operate in several distinct modes, each offering specific advantages, including harmonic amplification \citenumns{Zhao_2023}, cross-harmonic amplification \citenumns{Zhao_2023,zhao2025_nondegeneratepumping}, amplification with intrinsic pump-signal separation \citenumns{zhao2024_intrinsic_separation}, and phase-sensitive amplification \citenumns{zhao2025_nondegeneratepumping}. Translating these modes of operation to millimetre wavelengths is likely feasible and will enhance the versatility of the technology. In particular, a systematic study should be conducted to compare millimetre-wave amplification on a higher-order harmonic using a longer resonator with amplification at the same frequency on a lower-order harmonic using a shorter resonator. In the latter case, millimetre-wave ResPAs can be realised with much shorter transmission lines, enabling the fabrication of hundreds or even thousands of devices on a single wafer.

One particularly promising development of parametric amplifiers is their operation at the $3-4\,\mathrm{K}$ temperature range, which we have previously demonstrated in \citenumns{Zhao_2023,zhao2024_intrinsic_separation,zhao2025_nondegeneratepumping}. This temperature range can be maintained using the robust pulse-tube cooler technology which greatly reduces experimental complexity. For these applications, having a high $T_\mathrm{c}$ is likely advantageous as a reduced $T$ to $T_\mathrm{c}$ ratio will reduce the impact of loss and noise mechanisms associated with quasiparticle recombination \citenumns{Jonas_review}. Other studies on NbN deposition have shown that the $T_\mathrm{c}$ can be increased to above $15\,\mathrm{K}$ by performing additional optimisation against the argon flow rate, albeit possibly at the cost of lower resistivity \citenumns{NbN_high_Tc,NbN_hysteresis}. Further studies should be done to optimise NbN films for these high-temperature applications. 
 
\begin{acknowledgments}
The authors are grateful for funding from the UK Research and Innovation (UKRI) and the Science and Technology Facilities Council (STFC) through the Quantum Technologies for Fundamental Physics (QTFP) programme (Project Reference ST/T006307/2).

The authors are also grateful to Dr. Boon Kok Tan and Dr. Nikita Klimovich of the Superconducting Quantum Detectors Group at the University of Oxford for access to the dilution refrigerator and help with the measurements.
\end{acknowledgments}

\bibliographystyle{h-physrev}
\bibliography{library}
\end{document}